\documentclass[aps,english,prl,superscriptaddress,showpacs,twocolumn,nobalancelastpage]{revtex4-1}

\usepackage{amsfonts,amssymb,amsmath}
\usepackage{graphics,graphicx,epsfig}
\usepackage{amsthm}
\usepackage{amsmath}
\usepackage{epstopdf}
\usepackage{setspace}
\usepackage{booktabs}
\usepackage{bbold}
\usepackage{tabularx}
\usepackage{hyphenat}

\def\identity{\leavevmode\hbox{\small1\kern-3.8pt\normalsize1}}

\newcolumntype{L}[1]{>{\raggedright\arraybackslash}p{#1}}
\newcolumntype{C}[1]{>{\centering\arraybackslash}p{#1}}
\newcolumntype{R}[1]{>{\raggedleft\arraybackslash}p{#1}}

\newcommand{\ket}[1]{\left | #1 \right\rangle}

\newcommand{\Tr}{\mathrm{Tr}}

\renewcommand{\epsilon}{\varepsilon}

\makeatletter

\@ifundefined{textcolor}{}
{%
 \definecolor{BLACK}{gray}{0}
 \definecolor{WHITE}{gray}{1}
 \definecolor{RED}{rgb}{1,0,0}
 \definecolor{GREEN}{rgb}{0,1,0}
 \definecolor{BLUE}{rgb}{0,0,1}
 \definecolor{CYAN}{cmyk}{1,0,0,0}
 \definecolor{MAGENTA}{cmyk}{0,1,0,0}
 \definecolor{YELLOW}{cmyk}{0,0,1,0}
 }

\makeatother

\usepackage{babel}

\begin{document}

\renewcommand{\figurename}{FIG.}

\title{Experimental Comparison of Efficient Tomography Schemes for a Six-Qubit State}

\author{Christian~Schwemmer}
\affiliation{Max-Planck-Institut f\"ur Quantenoptik, Hans-Kopfermann-Stra{\ss}e 1, D-85748 Garching, Germany}
\affiliation{Department f\"ur Physik, Ludwig-Maximilians-Universit\"at, D-80797 M\"unchen, Germany}

\author{G\'eza~T\'oth}
\affiliation{Department of Theoretical Physics, University of the Basque Country UPV/EHU, P.O. Box 644, E-48080 Bilbao, Spain}
\affiliation{IKERBASQUE, Basque Foundation for Science, E-48011 Bilbao, Spain}
\affiliation{Wigner Research Centre for Physics, Hungarian Academy of Sciences, P.O. Box 49, H-1525 Budapest, Hungary}

\author{Alexander~Niggebaum}
\affiliation{School of Physics and Astronomy, University of Birmingham, B15 2TT Birmingham, United Kingdom}

\author{Tobias~Moroder}
\affiliation{Naturwissenschaftlich-Technische Fakult\"at, Universit\"at Siegen, Walter-Flex-Stra{\ss}e 3, D-57068 Siegen, Germany}

\author{David~Gross}
\affiliation{Physikalisches Institut, Universit\"at Freiburg \& FDM, Rheinstra{\ss}e 10, D-79104 Freiburg, Germany}

\author{Otfried~G\"uhne}
\affiliation{Naturwissenschaftlich-Technische Fakult\"at, Universit\"at Siegen, Walter-Flex-Stra{\ss}e 3, D-57068 Siegen, Germany}

\author{Harald~Weinfurter}
\affiliation{Max-Planck-Institut f\"ur Quantenoptik, Hans-Kopfermann-Stra{\ss}e 1, D-85748 Garching, Germany}
\affiliation{Department f\"ur Physik, Ludwig-Maximilians-Universit\"at, D-80797 M\"unchen, Germany}

\begin{abstract}
Quantum state tomography suffers from the measurement effort increasing exponentially with the number of qubits. Here, we demonstrate permutationally
invariant tomography for which, contrary to conventional tomography, all resources scale polynomially with the number of qubits both in terms of the 
measurement effort as well as the computational power needed to process and store the recorded data. We demonstrate the 
benefits of combining permutationally invariant tomography with compressed sensing by studying the influence of the pump power on the noise present 
in a six-qubit symmetric Dicke state, a case where full tomography is possible only for very high pump powers.
\end{abstract}


\maketitle

\emph{Introduction.---}The number of controllable qubits in quantum experiments is steadily growing \cite{MULTIQUBITS,MULTIQUBITS2}. Yet, to fully 
characterize a multiqubit state via quantum state tomography (QST), the measurement effort scales exponentially with the number of qubits. Moreover, 
the amount of data to be saved and the resources to process them scale exponentially, too. Thus, the limit of conventional QST will soon be reached. 
The question arises: how much information about a quantum state can be inferred without all the measurements a full QST would require. Protocols have 
been proposed which need significantly fewer measurement settings if one has additional knowledge about a state, e.g., that it is of
low rank, a matrix product state or a permutationally invariant (PI) state \cite{TOMOPAPERS,COMPRESSEDSENSING,TOMOPAPERS2,TOMOPAPERS3,PITOMO,TOBISFIT}.
Some of these approaches only require a polynomially increasing number of measurements and even offer scalable post-processing algorithms \cite{TOMOPAPERS2,TOBISFIT}.
Yet, it is important to test the different approaches and evaluate their results for various quantum states.

Here we implement and compare four different QST schemes in a six-photon experiment. In detail, we perform the largest QST of a photonic multiqubit 
state so far. We use these data as a reference for a detailed evaluation of different tomography schemes, which enable the state determination with 
significantly fewer measurements. The recently proposed, scalable PI analysis is implemented here and thus enables us, for the first time, to also perform 
the numerical evaluation with polynomial resources only. We evaluate the convergence of compressed sensing (CS) schemes and show that the combination of PI and CS can further 
reduce the measurement effort, without sacrificing performance. We demonstrate the usability of these significantly improved methods to characterize 
the effects of higher-order emission in spontaneous parametric down-conversion (SPDC), an analysis which would not have been possible without the novel 
tomography schemes.

%
%
%
%

\emph{Scalable scheme for measurements.---}Let us first consider the measurement effort needed for tomography. For full QST, each 
\nolinebreak{$N$-qubit} state is associated with a normalized non-negative Hermitian matrix $\varrho$ with $4^{N}-1$ real free parameters. Since all 
free parameters have to be determined, any scheme suitable to fully analyze an arbitrary state, such as, e.g., the standard Pauli tomography scheme, suffers 
from an exponentially increasing measurement effort \cite{QSESTIMATION, THOMOBASICS}. PI states in contrast are described by only 
${N+3 \choose N}-1 = O(N^3)$ free parameters. Tomography in the PI subspace can be performed by measuring (global) operators of the form 
$A_i^{\otimes N}$ with $A_i = \vec{n}_i\vec{\sigma}$, i.e., measurements of the polarization along the same direction $\vec{n}_i$ for every photon 
\cite{PITOMO}. Here, $|\vec{n}_i| = 1$ and $ \vec{\sigma} = (\sigma_x,\sigma_y,\sigma_z)$ with Pauli operators $\sigma_i$ ($i=x,y,z$). Each single 
measurement setting $A_i^{\otimes N}$ delivers $N$ expectation values of the operators 
$M_i^n = \frac{1}{N!}\sum_k \Pi_k [|0\rangle_i\langle0|^{\otimes(N-n)} \otimes |1\rangle_i\langle1|^{\otimes n}]\Pi_k^\dagger,$ where the summation 
is over all permutations $\Pi_k$ and $i$ refers to the eigenbasis of $A_i$. This reduces the number of necessary settings to 
$\mathcal{D}_{N} = {N + 2 \choose N} = \frac{1}{2}(N^2+3N+2)=O(N^2)$. Note, if one allows global entangled measurements this number can be further 
reduced~\cite{SANCHEZSOTO}. Most importantly, whether an unknown $N$-qubit state is close to being PI can be checked in advance by measuring the 
settings $\sigma_x^{\otimes N}, \sigma_y^{\otimes N}, \sigma_z^{\otimes N}$. These measurements are already sufficient to give a lower bound for 
the overlap with the symmetric subspace~\cite{PITOMO,PSFORMEL}.

%
%
%
%

\emph{Scalable representation of states and operators.---}The above approach not only reduces the experimental effort, it also offers the possibility 
to efficiently store and process the measured data. Describing states in the PI 
subspace enables an efficient representation with only polynomial scaling of the storage space and processing time \cite{TOBISFIT, STEINBERG}. 

\begin{figure}[!t]
\includegraphics[width=0.28\textwidth]{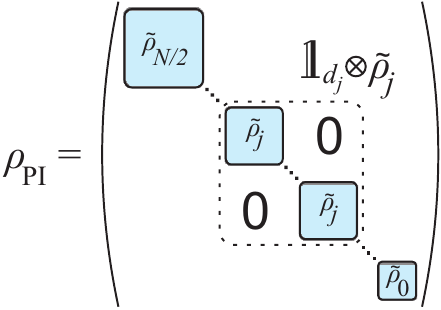}
\caption{(color online). Every PI state can be decomposed into a block diagonal form. 
Exemplarily shown is the combination of  $d_j$ block matrices $\widetilde{\varrho_j}$  which are all identical.}
\label{decomposition}
\end{figure}

Consider the angular momentum basis states $\vert j,j_z,\alpha\rangle$ for the $N$-qubit Hilbert space, with 
$\vec{J}^2\vert j,j_z,\alpha\rangle=j(j+1)\vert j,j_z,\alpha\rangle,$ $J_z\vert j,j_z,\alpha\rangle=j_z\vert j,j_z,\alpha\rangle,$ 
where the total spin numbers are restricted to be $j = j_{\min},j_{\min}+1,...,\frac{N}{2}$ starting from $j_{\min} = 0$ for $N$ even and 
$j_{\min} = \frac{1}{2}$ for $N$ odd, while $j_z=-\frac{N}{2},-\frac{N}{2}+1,...,\frac{N}{2}$. Here, $\alpha=1,2,...,d_j$ is a label to remove the 
degeneracy (of degree $d_j$~\cite{DEGENERACY}) of the eigenstates of $\vec{J}^2$ and $J_z^2$.  In this basis, PI states can be written in a simple block diagonal form 
\begin{eqnarray}
\varrho_{\textrm{PI}} = \bigoplus\limits_{j=j_{\min}}^{N/2} \frac{\identity_{d_j}}{d_j} \otimes p_j \varrho_j
\label{decompostioneq}
\end{eqnarray}
with $\varrho_j$ being the density operators of the spin-$j$ subspace and $p_j$ a probability distribution. Hence, it is sufficient to consider only 
the $\frac{N}{2}$ blocks $\widetilde{\varrho_j} = p_j \varrho_j  / d_j$ (of which each has a multiplicity of $d_j$; see Fig.~\ref{decomposition}) 
with the largest block --- the symmetric subspace --- being of dimension $(N+1) \times (N+1)$ and multiplicity $d_{\frac{N}{2}}=1$. Consequently, a 
PI state can be stored efficiently.

Even if the state to be analyzed is not PI, as long as the observable to be measured is PI one can hugely benefit from the scheme, since a similarly
scalable decomposition can be found for any PI operator $O$, i.e., $O = \bigoplus_j \identity_{d_j} \otimes O_{j}$. Together 
with Eq. (\ref{decompostioneq}) this yields an efficient way to also calculate the expectation values 
$\langle O \rangle = \Tr{(\varrho O)} = \sum_j p_j \Tr{(\varrho_j O_j)}$ for non-PI states.
Note that while, in the regular case, the trace has to be taken over the product of two $2^N$-dimensional matrices, now we only have about $\frac{N}{2}$ 
terms with traces of at most $(N+1)$-dimensional matrices. Again, the effort reduces from exponential to polynomial. For the six-qubit case 
($j \in j_{\textrm{min}} = 0, 1, 2, \frac{N}{2} = 3$) this means that the state to be analyzed as well as each measurement operator can be described 
by only four Hermitian matrices of size $7\times7$, $5\times5$, $3\times3$ and $1\times1,$ respectively, reducing the number of parameters from 
$4^6-1=4095$ to ${9 \choose 6}-1 = 83$ only.

%
%
%
%
%

Data analysis starts with the counts $c_i^n$ observed measuring $M_i^n$ and the frequencies $f_i^n = c_i^n/\sum_k c_i^k$, respectively. Solving the 
system of linear equations $f_i^n \approx \langle M_i^n \rangle = \Tr{(\varrho M_i^n)}$ for the free parameters of $\varrho$ usually results in a 
nonpositive and thus unphysical density matrix ($\varrho \ngeq 0$) due to statistical errors and  misalignment. Here, typically, a maximum likelihood 
(ML) fitting algorithm is used to find the physical state that optimally agrees with the measured data \cite{QSESTIMATION,HRADILML,SUPPLEMENT}.
We use convex optimization \cite{TOBISFIT,CONVEXOPTIMIZATION} which guarantees a unique minimum and fast convergence. The performance of 
our algorithm is illustrated best by the fact that a 20-qubit PI state can be reconstructed in less than 10\,min on a standard desktop computer.

%
%
%
%
%
%

\emph{State reconstruction of low rank states and compressed sensing.---}As shown recently, low rank states, i.e., states with only few non-zero 
eigenvalues, enable state reconstruction even if the underlying set of data obtained from random Pauli measurements is incomplete 
\cite{COMPRESSEDSENSING}. There, the measurement effort to analyze a state of rank $r$ with $r2^N$ free parameters scales like $O(r2^N \log 2^N)$ --
clearly achieving optimal scaling up to a log factor. Despite the still exponential scaling, the square root improvement can be considerable. 
Since, in many cases, the state to be experimentally prepared is at the same time PI \textit{and} of low rank, we demonstrate here for the first time that 
combining the two methods is possible \cite{lowrank,SUPPLEMENT}.

\emph{Experimental state tomography.---}Let us now compare the various QST schemes. In particular we evaluate 
the number of settings necessary to obtain (almost) full knowledge about the state. As a reference, we perform, for the first time, full QST of a six-photon 
state. This is possible only at very high pump power (8.4\,W) of the down-conversion source where we collect data for the complete set of Pauli settings. PI 
tomography is performed to test it against full QST and to analyze states emitted for lower pump powers. For both strategies, we also analyze the convergence
of CS tomography for incomplete data.

The six-photon state observed in this work is the symmetric Dicke state $\vert D_{6}^{(3)}  \rangle $. In general, symmetric Dicke states are 
defined as
\begin{equation}\label{DickeState}
\vert D_{N}^{(n)}  \rangle =\binom{N}{n}^{-1/2} \sum_i \mathcal{P}_i(\vert H^{\otimes(N-n)} \rangle \otimes  \vert V^{\otimes n} \rangle ),
\end{equation}
where $\ket{H/V}_i$ denotes horizontal or vertical polarization in the $i^{\textrm{th}}$ mode and the $\mathcal{P}_i$ represent all the distinct 
permutations.
%
%
%
%
%
In order to experimentally observe $\vert D_{6}^{(3)}  \rangle$, we distribute an equal number of $H$ and $V$ polarized photons over six output modes and 
apply conditional detection (fore details see the Supplemental Material~\cite{SUPPLEMENT} and \cite{DICKEPAPERS}). 
The setup uses cavity enhanced SPDC \cite{CAVITY} with special care taken to further reduce losses of all components 
and to optimize the yield of $|D_{6}^{(3)}\rangle$.
\begin{table}[!t]	
		\begin{tabular*}{86mm}{L{10mm}  C{18.5mm} C{18.5mm} C{18.5mm}  C{18.5mm}}
		\hline\hline
			State &  Full &  PI & CS & PI,CS \\\hline
			$|D_{6}^{(0)}\rangle$ & $0.001$ & $0.001$ & $0.001$ & $0.002$\\
			$|D_{6}^{(1)}\rangle$ & $0.005$ & $0.008$ & $0.011$ & $0.006$\\
			$|D_{6}^{(2)}\rangle$ & $0.197$ & $0.222$ & $0.181$ & $0.207$\\
			$|D_{6}^{(3)}\rangle$ & $0.604$ & $0.590$ & $0.615$ & $0.592$\\
			$|D_{6}^{(4)}\rangle$ & $0.122$ & $0.127$ & $0.118$ & $0.119$\\
			$|D_{6}^{(5)}\rangle$ & $0.003$ & $0.004$ & $0.003$ & $0.005$\\
			$|D_{6}^{(6)}\rangle$ & $0.000$ & $0.003$ & $0.001$ & $0.004$\\\hline
			$\sum$ & $0.933$ & $0.954$ & $0.929$ & $0.935$\\
			\hline\hline			
		\end{tabular*}
		\caption{Overlap with the symmetric Dicke states determined from full tomography, PI tomography with 28 settings, CS with 270 
		settings and CS in the PI subspace (PI,CS) with 16 settings. The fidelities for all tomography schemes were determined from the 
		respective ML reconstructed states. 		Nonparametric bootstrapping \cite{BOOTSTRAP} was performed from 
		which the corresponding standard deviations were determined as $<0.005$, $<0.015$, $<0.008$, and $<0.020$ for full tomography, PI 
		tomography, CS, and CS in the PI subspace, respectively.}
		\label{TABLE_FIDELITIES}
\end{table}

\begin{figure}[!t]
\includegraphics[width=0.45\textwidth]{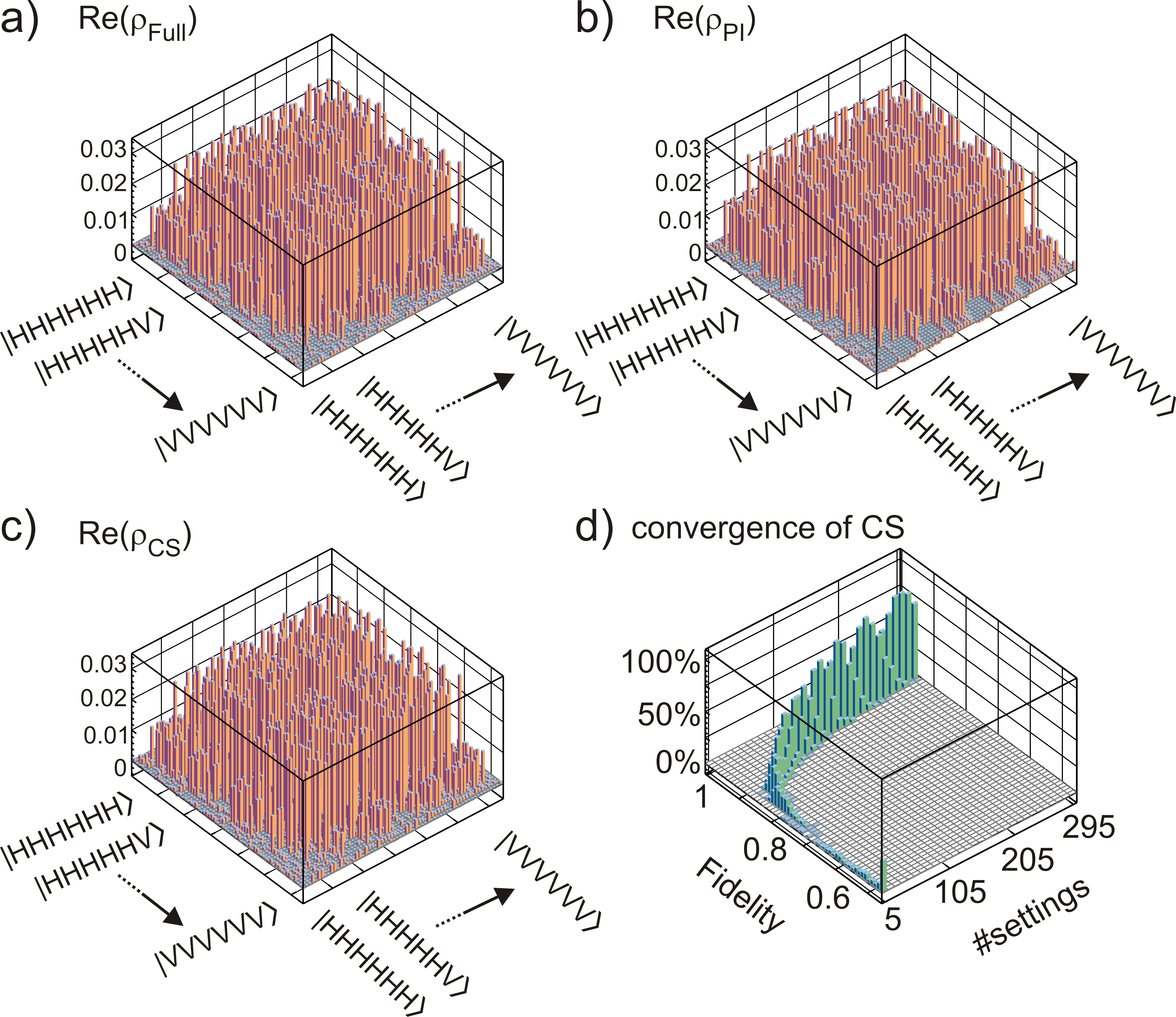}
\caption{(color online). ML reconstruction of the state $\vert D_{6}^{(3)}  \rangle$ obtained from (a) full (b) PI tomography and (c) CS with 270 
settings performed at a pump power of 8.4\,W. The respective fidelities are $0.604$, $0.590$ and $0.615$ with a mutual overlaps of 
$F(\varrho_{\textrm{full}},\varrho_{\textrm{PI}})=0.922$, $F(\varrho_{\textrm{full}},\varrho_{\textrm{CS}})=0.950$ and 
$F(\varrho_{\textrm{PI}},\varrho_{\textrm{CS}})=0.908$. (d) Probability to obtain a certain fidelity for CS with a certain number
of randomly chosen settings in comparison with full tomography.} 
\label{symmetric}
\begin{center}
\includegraphics[width=0.45\textwidth]{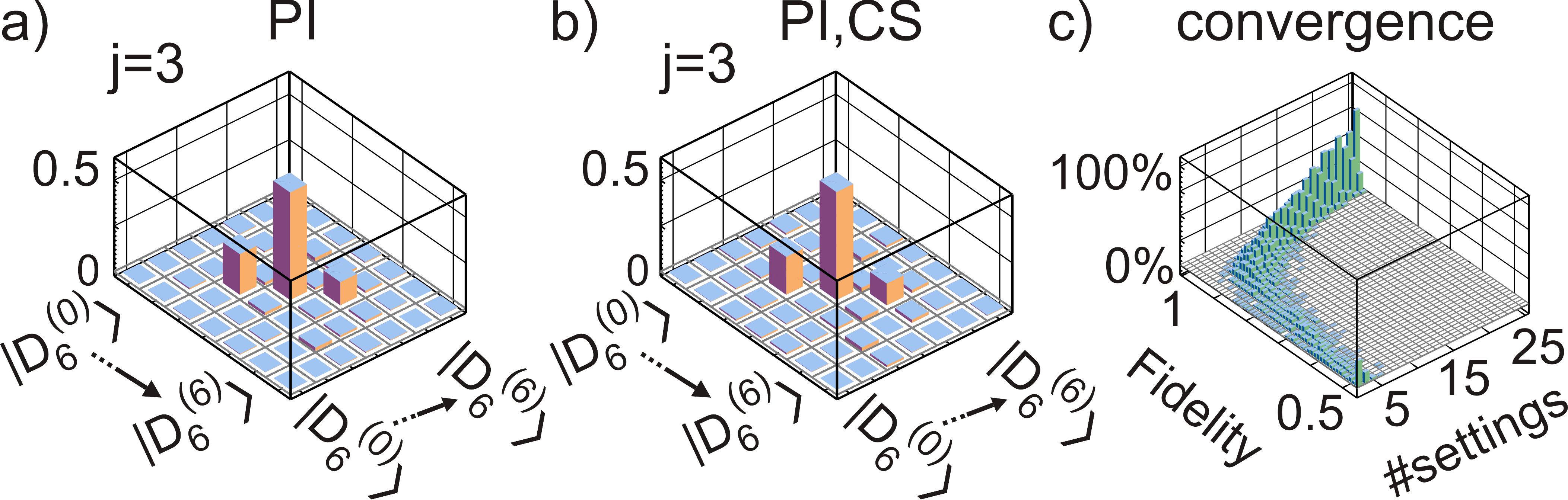}
\caption{(color online). Symmetric subspaces ($j=3$) obtained with (a) PI tomography and (b) CS in the PI subspace with 16 settings. The central 
bars can be associated with the target state $\vert D_{6}^{(3)}  \rangle$ and the small bars next to it with $\vert D_{6}^{(2)}  \rangle$ and 
$\vert D_{6}^{(4)}  \rangle$ originating from higher-order noise. (c) Probability to observe a certain fidelity for arbitrarily chosen 
tomographically incomplete sets of settings in comparison with PI tomography from 28 settings. For 16 settings the overlap is $\geq 0.950$ on average.}
\label{PICS_CS}
\end{center}
\end{figure}

Data are recorded at a pump power of $8.40 \pm 0.56$\,W over 4\,min for each of the $3^6 = 729$ Pauli settings. The six-photon count rate was 
58 events per minute on average, leading to about 230 events per basis setting within a total measurement time of approximately 
50~h \cite{runtime}. The reconstructed density matrix can be seen in Fig.~\ref{symmetric}(a). 
Table \ref{TABLE_FIDELITIES} lists the fidelity \cite{uhlmann} with all the various Dicke states. Their sum reaches high values proving that the 
state is close to the symmetric subspace.

Evidently, the experimental state is a mixture of mainly $|D_6^{(2)} \rangle$, $|D_6^{(3)} \rangle$, and $|D_6^{(4)} \rangle$, and thus CS might be 
used beneficially. The following question arises: how many settings are required for CS for a faithful reconstruction of the state? We chose random 
subsets of up to 300 settings from the 729 settings for full tomography. Figure~\ref{symmetric}(d) gives the probability distribution of the fidelity 
of the reconstructed matrix for a bin size of $0.01$ with respect to the results of full tomography. While, for a low number of settings ($<10$), the 
results are randomly spread out, the overlap is already, on average, $\geq 0.800$ for 20 settings. We find that to reach a fidelity of $\geq 0.950$, 
one requires about 270 settings. Figure~\ref{symmetric}(c) shows the density matrix obtained from 270 settings 
$[F(\varrho_{\textrm{CS}},\varrho_{\textrm{full}}) = 0.950]$.

PI tomography should be clearly more efficient. To test its applicability, we first determined the lower bound for the projection of the state 
onto the symmetric subspace, i.e., the largest block in Fig.~\ref{decomposition}, $\langle P_{\rm s}^{(6)} \rangle$ from the settings 
$\sigma_x^{\otimes 6}, \sigma_y^{\otimes 6}$, and $\sigma_z^{\otimes 6}$ by analyzing all photons under $\pm45^{\circ}$, right- or left-circular, 
and  $H/V$ polarization. We found $\langle P_{\rm s}^{(6)} \rangle \geq 0.922\pm0.055$, indicating that it is legitimate to use PI tomography, which 
for six qubits only requires 25 more settings \cite{SUPPLEMENT}. Under the same experimental conditions as before and 4\,min of data collection per setting,
we performed the experiment within 2\,h only. The density matrix $\varrho_{\textrm {PI}}$ obtained is shown in Fig.~\ref{symmetric}(b), with its symmetric 
subspace shown in Fig.~\ref{PICS_CS}(a). The fidelities with the symmetric Dicke states for PI tomography can be found again in 
Table \ref{TABLE_FIDELITIES}. (For the projector to the Dicke state $ |D_N^{(n)} \rangle$, all $\{O_{j}\}_{kl}=0$ except for 
$\{O_{\frac{N}{2}}\}_{n+1,n+1}=1$). The overlap between the reconstructed states using either full or PI tomography is $0.922$, which is equivalent 
to the fidelity of $0.923$ between full tomography and its PI part. Clearly, PI tomography rapidly and precisely determines the PI component of the 
state.

\emph{PI tomography with CS.---}To speed up analysis even further, based on subsets of the data used for PI tomography, we derived the density matrix 
$\varrho_{\rm PI,CS}$; see Fig. \ref{PICS_CS}(b). Here, the fidelity averaged over a series of different samples is above 0.950 for 16 or more 
settings [Fig. \ref{PICS_CS}(c)]. Again, both methods are compatible within 1 standard deviation. In summary, our results prove that PI 
tomography (with CS) enables precise state reconstruction with minimal experimental and computational effort. 

\begin{figure}[ht]
\includegraphics[width=0.40\textwidth]{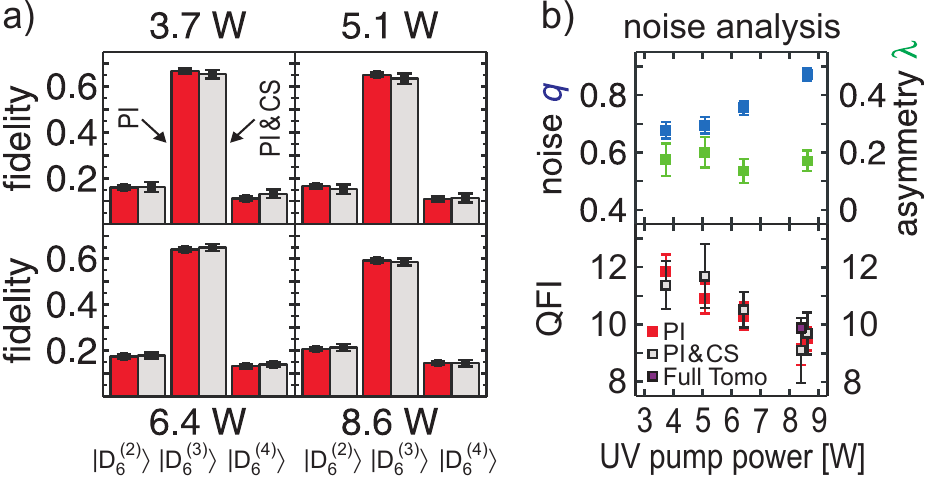}
\caption{(color online). (a) Observed fidelities with the states $\vert{D_6^{(2)}}\rangle$, $\vert{D_6^{(3)}}\rangle$ and $\vert{D_6^{(4)}}\rangle$ 
at different ultra violet (UV) pump powers for PI tomography and CS in the PI subspace from 12 settings. The error bars were determined by 
nonparametric bootstrapping \cite{BOOTSTRAP}. (b) The influence of the pump power on the higher-order noise expressed via the 
noise $q$ and the asymmetry parameter $\lambda$ (upper part) and the phase estimation sensitivity expressed via the quantum Fisher information (QFI) (lower part).}
\label{noise}
\end{figure}

\emph{Application to noise analysis.---}As the count rates for six-photon states depend on the cube of the pump power, full QST is not possible for 
lower pump power within reasonable time and thus does not allow to analyze the features of multiphoton states obtained 
form SPDC. As SPDC is a spontaneous process, with certain probability, there are cases where eight photons have been emitted but only six have 
been detected, leading to an admixture of $\varrho_{D_6^{(2)}}$ and $\varrho_{D_6^{(4)}}$. Ideally, the amplitude of the two admixtures 
should be the same, but due to polarization dependent coupling efficiencies of $H$ and $V$ photons \cite{COUPLING_1, COUPLING_2}, this is not the case. 
Therefore, we extended the noise model \cite{THESISWITLEF} to better specify the experimental state using 
$\varrho_{\mathrm{exp}}^{\mathrm{noise}}(q,\lambda) = (1-q) \varrho_{{D_6^{(3)}}} +q\varrho_6^{\mathrm{asym}}(\lambda)$, with 
$\varrho_6^{\mathrm{asym}}(\lambda) = \frac{4}{7}\varrho_{{D_6^{(3)}}} + \frac{3}{14}\left[(1+\lambda)\varrho_{{D_6^{(2)}}} + 
(1-\lambda)\varrho_{{D_6^{(4)}}}\right]$,  the noise $q$, and  the asymmetry parameter $\lambda$. Both $q$ and $\lambda$ can be determined from the 
fidelities to the Dicke states (see also the Supplemental Material~\cite{SUPPLEMENT}). At $8.4$\,W noise parameters of $q = 0.807 \pm 0.013$ and 
$\lambda = 0.234 \pm 0.015$ were obtained from full tomography, which agree well with those from PI tomography ($q = 0.867 \pm 0.041$ and 
$\lambda = 0.273 \pm 0.059$). After convincing ourselves that (CS) PI tomography is in excellent agreement with full QST, we now also perform 
tomography for low pump powers. 

We performed PI analysis at 3.7, 5.1, 6.4, and 8.6\,W [see Fig.~\ref{noise}(a)] with sampling times of 67, 32, 18, 
15\,h and average counts per setting of 340, 390, 510, and 610, respectively.
PI tomography shows an increase of the noise parameter $q$ from $0.677 \pm 0.029$ for 3.7\,W to $0.872 \pm 0.023$ for 8.6\,W due to the increasing 
probability of eight-photon emission for high pump power~\cite{CSRESULTS}. Note, the ratio between six-photon detection from eight-photon emission relative to 
detection from six-photon emission is given by  $q/(1-q)$, i.e., for a pump power of 8.6\,W, we obtain sixfold detection events with 90\% probability 
from eight-photon emissions, of which two photons were lost. Although fluctuating, the asymmetry parameter $\lambda$ does not show significant 
dependence on the pump power  and lies in the interval $[0.136 \pm 0.042, 0.200 \pm 0.053]$ for PI tomography (within 
[$0.101 \pm 0.116, 0.190 \pm 0.071$] for CS in the PI subspace). This confirms that the difference in the coupling efficiency of $H$ and $V$ does not 
change with the pump power [see Fig.~\ref{noise}(b)]. The fidelity between the ML fits and the noise model 
$\varrho_{\mathrm{exp}}^{\mathrm{noise}}(p,\lambda)$ is $>0.925$ for all pump levels, and, for CS in the PI subspace, it is $>0.897$. The high values 
indicate that our noise model adequately describes the experimental results.    

As an example where full knowledge of $\varrho$ is necessary, let us consider the quantum Fisher information $F_Q$ which
measures the suitability of $\varrho$ to estimate the phase $\theta$ in an evolution $U(\theta,{\cal H}) = e^{-i \theta {\cal H}}$~\cite{FISHER}.
Here, we want to test whether, in spite of the higher-order noise, the reconstructed states still exhibit sub-shot-noise
phase sensitivity. 
For ${\cal H}$ we choose the collective spin operator $J_x = \sum_{i=1}^{N} \sigma_x^{(i)}$, where $\sigma_x^{(i)}$ is $\sigma_x$ acting on the i$th$ particle. 
In the case $N=6$, a value $F_Q>6$ indicates sub-shot-noise phase sensitivity. We observed
$11.858 \pm 0.576$, $10.904 \pm 0.528$, $10.289 \pm 0.468$, $9.507 \pm 0.411$ for the corresponding pump powers from 3.7\,W to 8.6\,W~\cite{CSRESULTS} [see Fig.~\ref{noise}(b)];
i.e., sub-shot-noise phase sensitivity is maintained for high pump powers.

\emph{Conclusions.---}We compared standard quantum state tomography with the significantly more efficient permutationally invariant tomography and 
also with compressed sensing in the permutationally invariant subspace. For this purpose, we used data of the symmetric Dicke state 
$\vert{D_6^{(3)}}\rangle$ obtained from spontaneous parametric down-conversion of very high pump power. All methods give compatible results within their statistical errors. 
The number of measurement settings was gradually reduced from 729 for full tomography, to 270 for compressed sensing, to 28 for permutationally 
invariant tomography, and to only 16 for compressed sensing in the permutationally invariant subspace, giving, in total, a reduction of about a factor of 
50 without significantly changing the quantities specifying the state. We applied this highly efficient state reconstruction scheme to study the 
dependence of higher-order noise on the pump power, clearly demonstrating its benefits for the analysis of 
multiqubit states required for future quantum computation and quantum simulation applications.

\emph{Acknowledgments.---}We thank R. Krischek, W. \nohyphens{Wieczorek}, Z. Zimbor\'as, S. Neuhaus and L. Knips for stimulating discussions. 
We acknowledge the support of this work by the EU (QWAD, ERC StG GEDENTQOPT, ERC QOLAPS, CHIST-ERA QUASAR, Marie Curie CIG 293993/ENFOQI), the Excellence 
Initiative of the German Federal and State Governments (ZUK 43), the DFG, FQXi Fund (Silicon Valley Community Foundation), the MINECO (Project No. FIS2012-36673-C03-03), the Basque Government (Project No.
IT4720-10), and the National Research Fund of Hungary OTKA (Contract No. K83858). 
D. Gross acknowledges grants W911NF-14-1-0098 and W911NF-14-1-0133 from the ARO. C. S. thanks QCCC of the Elite Network of Bavaria for support.

\makeatletter
\renewcommand{\@biblabel}[1]{[#1]} 
\makeatother

\appendix

%
%
\renewcommand{\thefigure}{S\arabic{figure}}
\renewcommand{\thetable}{S\arabic{table}}
\renewcommand{\theequation}{S\arabic{equation}}
\setcounter{figure}{0}
\setcounter{table}{0}
\setcounter{equation}{0}
\setcounter{section}{0}
\setcounter{page}{1}
\newcommand{\loo}{{\lambda}}

\newcommand{\proofend}{\hfill\fbox\\\medskip }

\section{Supplemental Material}
\section{The setup}

\begin{figure}[!t]
\includegraphics[width=0.48\textwidth]{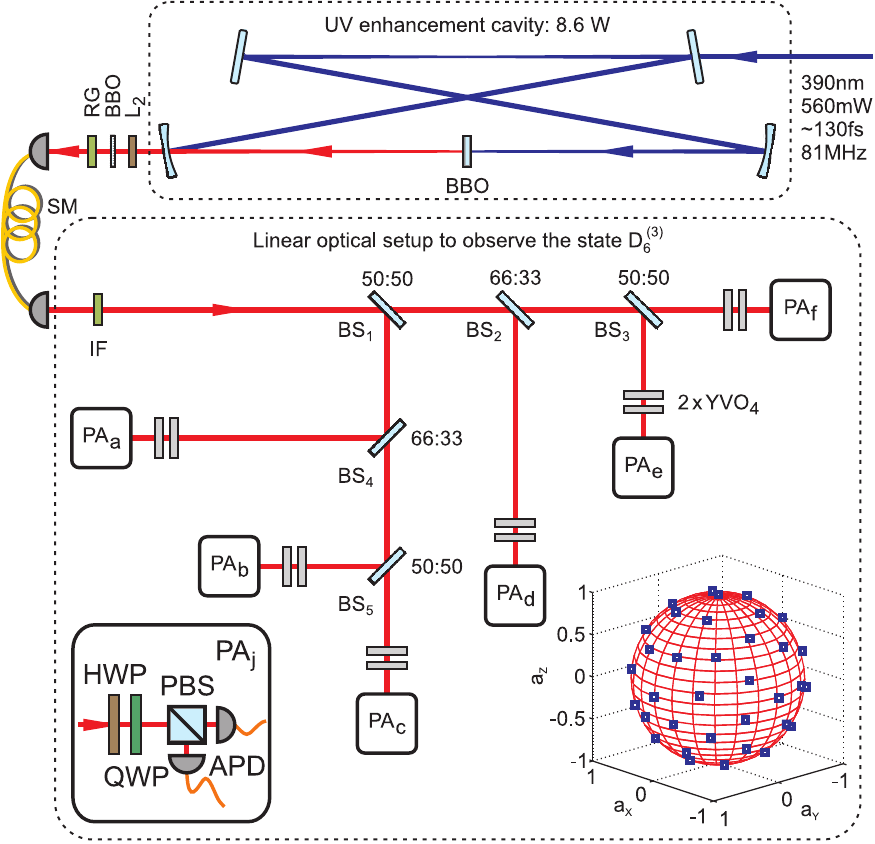}
\caption{Schematic drawing of the experimental setup to observe the symmetric Dicke state $\vert D_{6}^{(3)}  \rangle$ For a description, see text.}
\label{setup}
\end{figure}

The photon source is based on a femtosecond enhancement cavity in the UV with a 1\,mm thick $\beta$-barium-borate (BBO) crystal cut for type II phase 
matching placed inside \cite{CAVITY} [Fig. \ref{setup}]. In order to compensate for walk off effects a half-wave plate (HWP) and a second BBO crystal 
of 0.5\,mm are applied. Spatial  filtering is achieved by coupling the photons into a single mode fiber (SM) and an interference filter (IF) 
($\Delta \lambda$ = 3\,nm) enables spectral filtering. Distributing the photons into six spatial modes is realized by 3 beam splitters with a 
splitting ratio of 50:50 (BS$_1$, BS$_3$, BS$_4$) and two beam splitters with a ratio of 66:33 (BS$_2$, BS$_4$). Yttrium-vanadate (YVO$_4$) crystals 
are used to compensate for unwanted phase shifts. State analysis is realized by half-wave and quarter-wave plates (QWP) and polarizing beam splitters 
(PBS). The photons are detected by fiber-coupled single photon counting modules connected to a FPGA-based coincidence logic.

In Fig. \ref{setup} (lower right corner) a visualization of the measurement directions on the Bloch sphere is depicted. Each point $(a_x, a_y, a_z)$ 
on the sphere corresponds to a measurement operator of the form $a_x \sigma_x + a_y \sigma_y + a_z \sigma_z$. In order to perform PI tomography for 
six qubits  28 operators have to be measured.

\section{State reconstruction}
The target function to be minimized is the \emph{logarithmic likelihood} which is given by $\sum_{k,s} \frac{n_{k,s}}{N_{\rm{max}}}\log(p_{k,s})$ 
where $n_{k,s}$ labels the number of counts for the outcome $k$ when measuring setting $s$ with the corresponding probability $p_{k,s}$ for the guess 
$\hat \varrho$. In order to take into account slightly different total count numbers per setting, the $n_{k,s}$ have to be divided by the maximum 
count number observed in one setting $N_{\rm{max}} = {\rm max}(N_s)$.

For CS exactly the same target function has to be minimized with the only difference that the underlying set of measurement data is tomographically 
incomplete.

\section{Convergence of CS in the PI subspace}
\begin{figure}[!t]
\begin{center}
\includegraphics[width=0.48\textwidth]{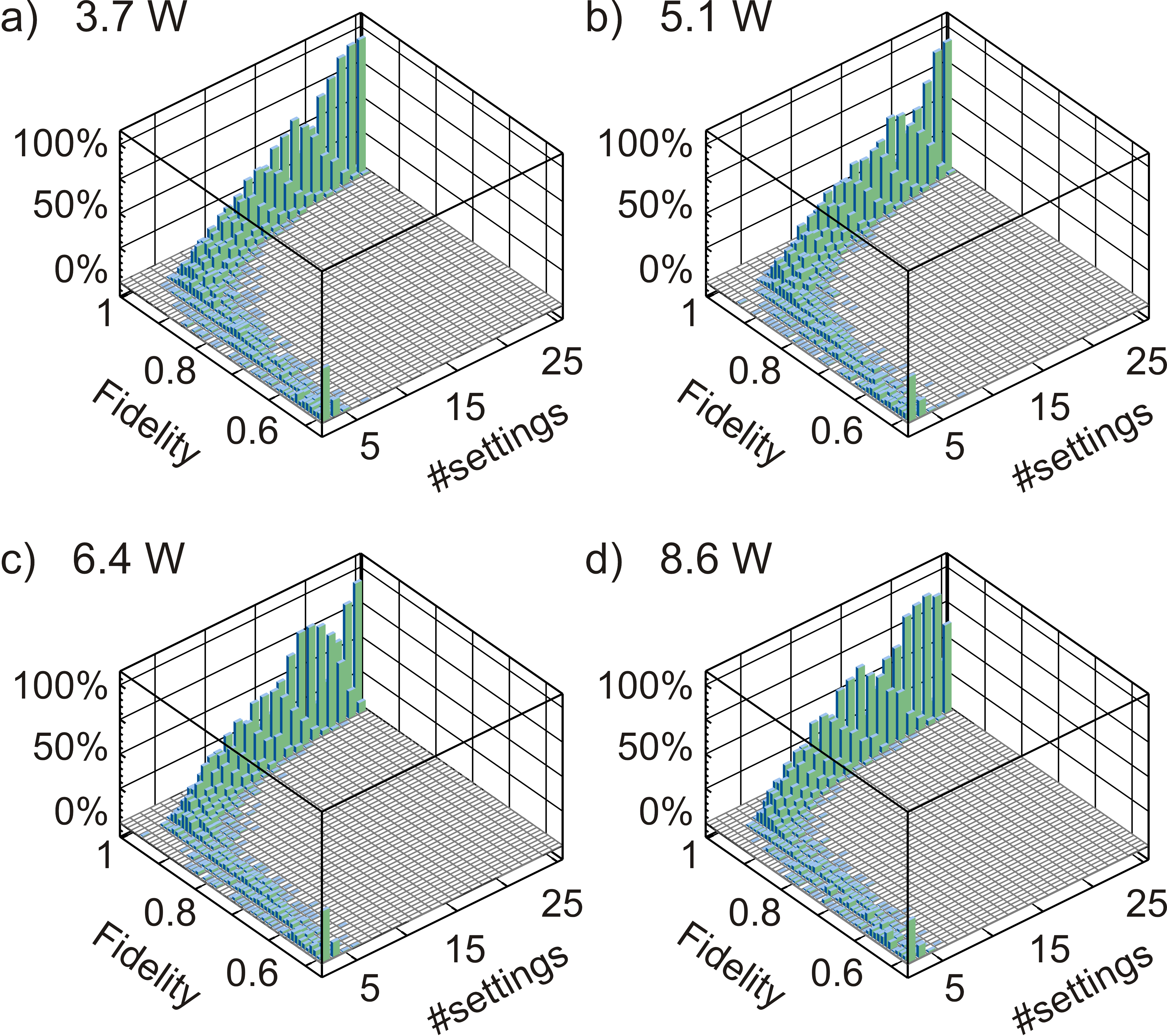}
\caption{Probability to observe a certain fidelity for arbitrarily chosen tomographically incomplete sets of settings in comparison with PI 
tomography from 28 settings at different pump levels. As soon as the number of settings surpasses 12, the state is almost perfectly determined, i.e., 
the overlap with respect to the states reconstructed from all settings $\geq 0.950$.}
\label{PICS}
\end{center}
\end{figure}

As described in the main text, we performed PI tomography together with CS in the PI subspace at different UV pump powers. In order to investigate 
the convergence of CS, series of different samples were randomly chosen from the full set of measurements. For all pump powers, the average fidelity 
with respect to all PI settings is above 0.950 as soon as the number of settings is $\geq 12$ (out of 28), see Fig. \ref{PICS}.

\section{Noise model}

\begin{figure}[!t]
\begin{center}
\includegraphics[width=0.40\textwidth]{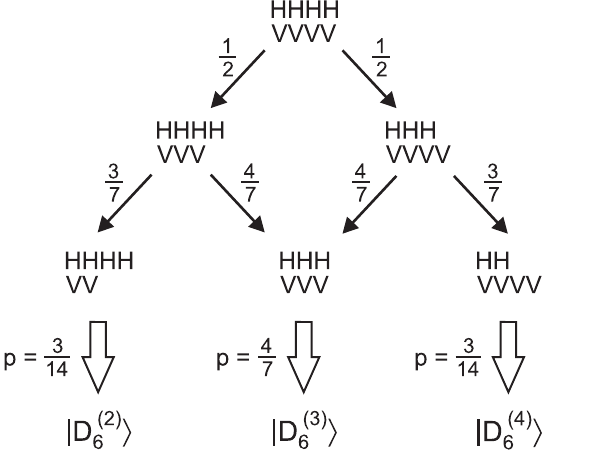}
\caption{The loss of two photons in an eight-photon event leads to an admixture of the state $\varrho_{D_6^{(2)}}$ and $\varrho_{D_6^{(4)}}$ to the target 
state. The respective probabilities $p$ can be determined by simple combinatorics.}
\label{HONOISE}
\end{center}
\end{figure}

As already explained in the main part of this paper, SPDC is a spontaneous process and therefore with a certain probability eight photons are emitted 
from the source. The loss of two of these eight photons in the linear optical setup and subsequent detection leads to an admixture of the states 
$\varrho_{D_6^{(2)}}$ and $\varrho_{D_6^{(4)}}$ for the case that either two $H$ or two $V$ polarized photons are not detected, respectively. 
However, in the case that one $H$ and one $V$ polarized photon remain undetected a considerable amount of this higher-order noise consists of the 
target state $\varrho_{D_6^{(3)}}$ thus preserving genuine multipartite entanglement even at high UV pump powers. The probabilities of the respective 
states to occur can be deduced from simple combinatorics, see Fig.~\ref{HONOISE}.
From this simple noise model, an experimental state of the form
\begin{equation}
\varrho_{\mathrm{exp}}^{\mathrm{noise}}(q,\lambda) = (1-q) \varrho_{{D_6^{(3)}}} +q\varrho_6
\end{equation}
with
\begin{equation}
\varrho_6 = \frac{4}{7}\varrho_{{D_6^{(3)}}} + \frac{3}{14}\left[\varrho_{{D_6^{(2)}}} + \varrho_{{D_6^{(4)}}}\right]
\end{equation}
would be expected. However, this is not observed experimentally since the emission angles of down-conversion photons are polarization dependent 
\cite{COUPLING_1,COUPLING_2} leading to an asymmetry in the coupling into the single mode fiber used. Therefore, the noisemodel was extended by the 
asymmetry parameter $\lambda$. Both $q$ and $\lambda$ can be deduced form the fidelities $F$ with respect to the Dicke states 
$\vert{D_6^{(2)}}\rangle$, $\vert{D_6^{(3)}}\rangle$ and $\vert{D_6^{(4)}}\rangle$
\begin{eqnarray}
q &=& \frac{7}{3} \cdot \frac{F_{\vert{D_6^{(2)}}\rangle} + F_{\vert{D_6^{(4)}}\rangle}}{F_{\vert{D_6^{(2)}}\rangle} + F_{\vert{D_6^{(3)}}\rangle} 
+ F_{\vert{D_6^{(4)}}\rangle}},\nonumber\\
\lambda &=& \frac{F_{\vert{D_6^{(2)}}\rangle} - F_{\vert{D_6^{(4)}}\rangle}}{F_{\vert{D_6^{(2)}}\rangle} + F_{\vert{D_6^{(4)}}\rangle}}.
\end{eqnarray}

\section{Entanglement witness}

Entanglement witnesses with respect to symmetric states are PI operators and thus can be determined efficiently.
For detecting genuine multipartite entanglement, we used the entanglement witness
\begin{eqnarray}
 {\cal{W}} &=& 0.420 \cdot \openone -0.700 |D_6^{(3)} \rangle \langle D_6^{(3)} |  \\ \nonumber
           &-& 0.160 |D_6^{(2)} \rangle \langle D_6^{(2)} | -  0.140 |D_6^{(4)} \rangle \langle D_6^{(4)} |,
\label{eqn:WITNESS}
\end{eqnarray}
where an expectation value $\langle \cal{W} \rangle <$~0 rules out any biseparability.
In order to obtain $\cal{W}$ we take an operator of the form
\begin{eqnarray}
A_{\alpha}&=&\alpha\vert D_{6}^{(3)}\rangle\langle
D_{6}^{(3)}\vert+\beta\vert D_{6}^{(2)}\rangle\langle
D_{6}^{(2)}\vert \\ \nonumber
&+&(1-\alpha-\beta)\vert D_{6}^{(4)}\rangle\langle
D_{6}^{(4)}\vert.
\end{eqnarray}
An entanglement witness can be obtained as
\begin{equation}
 W_{\alpha}=\max_{PPT}\langle A_{\alpha}\rangle \cdot \openone-A_{\alpha}
\end{equation}
where the maximum for bipartite PPT states can be obtained with semidefinite programming~\cite{TW09}. For $\alpha=0.700,\beta=0.160$ we have for PPT states 
over all partitions $\max_{PPT}\langle A_{\alpha}\rangle=0.420$. It is important that semidefinite programming always finds the global optimum. A 
systematic generalization to construct witnesses for Dicke states can be found in~Ref.~\cite{LEONARDO_BERGMANN}.

Here, we want use this witness to test whether, in spite of the higher-order noise, the observed states are still genuinely six-partite entangled.
For the corresponding pump powers from 3.7\,W to 8.6\,W, we determined the expectation value of ${\cal{W}}$ as  $-0.088 \pm 0.006$, $-0.078 \pm 0.006$, $-0.075 \pm 0.006$ 
and $-0.048 \pm 0.005$ for PI tomography and $-0.082 \pm 0.011$, $-0.064 \pm 0.013$, $-0.083 \pm 0.009$ and $-0.044 \pm 0.009$ for CS in the PI subspace. 
Clearly, due to the high probability of $\varrho_{D_6^{(3)}}$ states in the higher-order noise the entanglement is maintained also for high pump 
powers.


\begin{thebibliography}{99}

\setlength{\itemindent}{.0cm}
\setlength{\labelwidth}{2.5cm} 


\bibitem{MULTIQUBITS}
H. H\"affner, W. H\"ansel, C. F. Roos, J. Benhelm, D. Chek-al-kar, M. Chwalla, T. K\"orber, U. D. Rapol, M. Riebe, P. O. Schmidt, C. Becher, 
O. G\"uhne, W. D\"ur and, R. Blatt, Nature (London) \textbf{438}, 643 (2005); 
%
T. Monz, P. Schindler, J. T. Barreiro, M. Chwalla, D. Nigg, W. A. Coish, M. Harlander, W. H\"ansel, M. Hennrich, and R. Blatt, Phys. Rev. Lett. 
\textbf{106}, 130506 (2011). 

\bibitem{MULTIQUBITS2}%
Y.-F. Huang, B.-H. Liu,	L. Peng, Y.-H. Li, L. Li, C.-F. Li, and G.-C. Guo, Nat. Commun. \textbf{2}, 546 (2011); 
X.-C. Yao, T.-X. Wang, P. Xu, H. Lu, G.-S. Pan, X.-H. Bao, C.-Z. Peng, C.-Y. Lu, Y.-A. Chen, and J.-W. Pan, Nat. Photonics
\textbf{6}, 225 (2012). 
%


\bibitem{TOMOPAPERS}
D. Gross, Y.-K. Liu, S. T. Flammia, S. Becker, and J. Eisert, Phys. Rev. Lett. \textbf{105}, 150401 (2010);
S. T. Flammia, D. Gross, Y.-K. Liu, and J. Eisert,  New J. Phys. \textbf{14}, 095022 (2012).
M. Ohliger, V. Nesme, D. Gross, Y.-K. Liu, and J. Eisert, arXiv:1111.0853;
M. Gu\c{t}\u{a}, T. Kypraios, and I. Dryden,  New J. Phys. {\bf 14}, 105002 (2012);
M. Ohliger, V. Nesme, and J. Eisert, New J. Phys. {\bf 15}, 015024 (2013);
A. Smith, C.A. Riofr\'{\i}o, B. E. Anderson, H. Sosa-Martinez, I. H. Deutsch, and P. S. Jessen, Phys. Rev. A {\bf 87}, 030102(R) (2013).

\bibitem{COMPRESSEDSENSING}
M. Cramer, M. B. Plenio, S. T. Flammia, R. Somma, D. Gross, S. D. Bartlett, O. Landon-Cardinal, D. Poulin, and Y.-K. Liu, Nat. Commun. \textbf{1}, 
149 (2010);
O. Landon-Cardinal and D. Poulin, New J. Phys. {\bf 14}, 085004 (2012);

\bibitem{TOMOPAPERS2}
T. Baumgratz, D. Gross, M. Cramer, and M. B. Plenio, Phys. Rev. Lett. \textbf{111}, 020401 (2013);
T. Baumgratz, A. N\"u\ss{}eler, M. Cramer, and M. B. Plenio, New J. Phys. {\bf 15}, 125004 (2013).

\bibitem{TOMOPAPERS3} J.O.S. Yin and S. J. van Enk, Phys. Rev. A \textbf{83}, 062110 (2011).

\bibitem{PITOMO}
G. T\'oth, W. Wieczorek, D. Gross, R. Krischek, C. Schwemmer, and H. Weinfurter, Phys. Rev. Lett. \textbf{105}, 250403 (2010).

\bibitem{TOBISFIT}
T. Moroder, P. Hyllus, G. T\'oth, C. Schwemmer, A. Niggebaum, S. Gaile, O. G\"uhne, and H. Weinfurter, New J. Phys. \textbf{14}, 105001 (2012).

\bibitem{QSESTIMATION}
M. Paris and J. \v{R}eh\'{a}\v{c}ek, {\it Quantum State Estimation } (Springer-Verlag, Berlin, Heidelberg, 2004).


\bibitem{THOMOBASICS}
N. Kiesel, Ph.D. thesis, Ludwig-Maximilians-Universit\"at M\"unchen (2007);
%
D. F. V. James, P. G. Kwiat, W. J. Munro, and A. G. White, Phys. Rev. A \textbf{64}, 052312 (2001).

\bibitem{SANCHEZSOTO}
A. B. Klimov, G. Bj\"ork, and L. L. S\'anchez-Soto, Phys. Rev. A \textbf{87}, 012109 (2013).

\bibitem{PSFORMEL}
For six qubits, we have
$P_{\rm s}^{(6)} \geq \frac{2}{225}(J_x^2 + J_y^2 + J_z^2) -\frac{1}{90}(J_x^4 + J_y^4 + J_z^4) + \frac{1}{450}(J_x^6 + J_y^6 + J_z^6)$
with $J_{i} = \frac{1}{2}\sum_k \sigma_i^{(k)}$ and $\sigma_i^{(k)}$ the application of $\sigma_i$ on the $k^{\textrm {th}}$ qubit.

\bibitem{STEINBERG}
R. B. A. Adamson, P. S. Turner, M. W. Mitchell, and A. M. Steinberg, Phys. Rev. A, \textbf{78} 033832 (2008).


\bibitem{DEGENERACY} 
J. I. Cirac, A. K. Ekert, and C. Macchiavello, Phys. Rev. Lett. {\bf 82}, 4344 (1999).


\bibitem{HRADILML}
Z. Hradil, Phys. Rev. A \textbf{55}, R1561 (1997).


\bibitem{SUPPLEMENT}
See Supplemental Material for additional derivations, which includes Ref.~\cite{TW09, LEONARDO_BERGMANN}.

\bibitem{TW09} G. T\'oth, W. Wieczorek, R. Krischek, N. Kiesel, P.
Michelberger, and H. Weinfurter, New J. Phys. {\bf 11},
083002 (2009).

\bibitem{LEONARDO_BERGMANN}
For efficient entanglement detection methods for PI density matrices,  see L. Novo, T. Moroder and O. G\"uhne,  Phys. Rev. A {\bf88}, 012305 (2013);
M. Bergmann and O. G\"uhne, J. Phys. A: Math. Theor. {\bf 46}, 385304 (2013).




\bibitem{CONVEXOPTIMIZATION}
S. Boyd and S. Vandenberghe, \textit{Convex Optimization} (Cambridge University Press, Cambridge, England, 2004).


\bibitem{lowrank} 
In order to apply CS and PI tomography together, it suffices that the $\rho_j$ matrices in Eq. (\ref{decompostioneq}) are of low rank. This also covers the case of
global density matrices of comparatively high rank, such as the PI multiqubit singlet state discussed in I. Urizar-Lanz, P. Hyllus, I. L. Egusquiza, M. W. Mitchell, and
G. T\'oth, Phys. Rev. A {\bf 88}, 013626 (2013). For this highly mixed state, for even $N,$ $\rho_0=1,$ $p_0=1,$ and all other $p_j$'s are zero.



\bibitem{DICKEPAPERS}
W. Wieczorek, R. Krischek, N. Kiesel, P. Michelberger, G. T\'oth, and H. Weinfurter, Phys. Rev. Lett. \textbf{103}, 020504 (2009);
%
R. Prevedel, G. Cronenberg, M. S. Tame, M. Paternostro, P. Walther, M. S. Kim, and A. Zeilinger, Phys. Rev. Lett. \textbf{103}, 020503 (2009).

\bibitem{CAVITY}
R. Krischek, W. Wieczorek, A. Ozawa, N. Kiesel, P. Michelberger, T. Udem, and H. Weinfurter, Nat. Photonics \textbf{4}, 170 (2010).

\bibitem{BOOTSTRAP}
B. Efron and R. J. Tibshirani, An introduction to the bootstrap (Chapman \& Hall, London, 1994).
%
The error bars computed via this commonly employed method are a mere quantitative estimate about the fluctuations of the reconstructed state. One should keep in mind 
that the state reconstruction unavoidably induces systematic errors, see C. Schwemmer, L. Knips, D. Richart, T. Moroder, M. Kleinmann, O. G\"uhne and H. Weinfurter, 
arXiv:1310.8465, which are not recognizable with bootstrap analysis.


\bibitem{runtime}
The overall runtime of the experiment was almost two weeks from which 50\,h of useful data could be extracted.

\bibitem{uhlmann}
We use the Uhlmann fidelity $F(\varrho_1,\varrho_2)= \Tr{ (\sqrt{\sqrt{\varrho_1}\varrho_2\sqrt{\varrho_1}})}^2$ between the states $\varrho_1$ and 
$\varrho_2$ which simplifies to $F(\varrho_1,\varrho_2)=\Tr{(\varrho_1 \varrho_2)}$ if one of the two states is pure.


\bibitem{COUPLING_1}
R. S. Bennink, Y. Liu, D. D. Earl, and W. P. Grice, Phys. Rev. A \textbf{74},  023802 (2006).


\bibitem{COUPLING_2}
P. Trojek, Ph.D. thesis, Ludwig-Maximilians-Universit\"at M\"unchen (2007).


\bibitem{THESISWITLEF}
W. Wieczorek, Ph.D. thesis, Ludwig-Maximilians-Universit\"at M\"unchen (2009).



\bibitem{CSRESULTS}
The corresponding noise parameter $q$ determined from CS in the PI subspace with 12 randomly chosen settings is $0.725 \pm 0.052$ for 3.7\,W and 
$0.886 \pm 0.040$ for 8.6\,W. \\
The expectation values for the Fischer information are  $11.371 \pm 0.842$, $11.685 \pm 1.113$, $10.500 \pm 0.613$, $9.688 \pm 0.733$ for CS in the PI subspace.


\bibitem{FISHER}
The Fisher information is defined as $F_Q(\theta,{\cal H}) = 2 \sum_{i,j} \frac{(\lambda_i - \lambda_j)^2}{\lambda_i + \lambda_j}|\langle i|{\cal H}|j\rangle|^2$ with $\{\lambda_i, |i\rangle\}$ 
the eigenspectrum of $\varrho$, see
P. Hyllus, W. Laskowski, R. Krischek, C. Schwemmer, W. Wieczorek, H. Weinfurter, L. Pezz\'e, and A. Smerzi, Phys. Rev. A \textbf{85}, 022321 (2012).



\end{thebibliography}
\end{document}